\documentclass[twocolumn,prl,10pt,showpacs,superscriptaddress]{revtex4-1}
\usepackage{graphicx}  
\usepackage{amssymb} 
\usepackage{natbib}
\usepackage{amsmath}
\usepackage{enumerate}
\usepackage{textcomp}
\usepackage{times}
\usepackage{mathtools}
\usepackage{epsfig}
\usepackage[colorlinks,linkcolor=blue,citecolor=blue,urlcolor=blue]{hyperref}
\usepackage{color}
\usepackage{graphicx}
\usepackage{tabularx}
\usepackage{soul}
\usepackage[mathscr]{euscript}
\usepackage{amssymb}
\usepackage[normalem]{ulem}
\usepackage{mathrsfs}
\usepackage{amsmath}
\usepackage{color}
\usepackage{bbold}
\usepackage{comment}
\usepackage{bm}
\usepackage{braket}
\usepackage{dsfont}
\usepackage{textcomp}
\usepackage{lipsum}
\usepackage{amsmath,amssymb}
\newcommand{\pt}{\partial}
\newcommand{\mb}{\mathbf}

\newcommand{\mc}{\mathcal}
\newcommand{\tr}{\tilde{\mathbf{r}}}
\newcommand{\chid}{\chi^{\dagger}}
\newcommand{\wext}{\omega_{\mbox{\scriptsize{ext}}}}
\newcommand{\wextt}{\tilde{\omega}_{\mbox{\scriptsize{ext}}}}
\newcommand{\Hext}{H_{\mbox{\scriptsize{ext}}}}

\newcommand{\egap}{\varepsilon_{\mbox{\tiny{gap}}}}

\newcommand{\Pbs}{\Psi^{\scalebox{0.7}{bs}}}
\newcommand{\ems}{\varepsilon_{\mbox{\tiny{MS}}}}

\newcommand{\wres}{\omega_{\mbox{\scriptsize{res}}}}

\newcommand{\Fext}{F_{\scalebox{0.7}{ext}}}

\newcommand{\Wconf}{W_{\scalebox{0.7}{conf}}}
\newcommand{\Vconf}{V_{\scalebox{0.7}{conf}}}

\newcommand{\Req}{R_x^{\scalebox{0.7}{eq}}}
\begin{document}

\title{Skyrmions Driven by Intrinsic Magnons}

\author{Christina Psaroudaki}
\affiliation{Department of Physics, University of Basel, Klingelbergstrasse 82, 4056 Basel, Switzerland}
\author{Daniel Loss}
\affiliation{Department of Physics, University of Basel, Klingelbergstrasse 82, 4056 Basel, Switzerland}

\date{\today}
\begin{abstract}
We study the dynamics of a skyrmion in a magnetic insulating nanowire in the presence of  time-dependent oscillating magnetic field gradients. These ac fields act as a net driving force on the skyrmion via its own intrinsic magnetic excitations. In a microscopic quantum field theory approach we include the unavoidable coupling of the external field to the magnons, which gives rise to time-dependent dissipation for the skyrmion. We demonstrate that the magnetic ac field induces a super-Ohmic to Ohmic crossover behavior for the skyrmion dissipation kernels with time-dependent Ohmic terms. The ac driving of the magnon bath at resonance results in a unidirectional helical  propagation of the skyrmion in addition to the otherwise periodic bounded motion.  \end{abstract}

\maketitle

\emph{Introduction.} The understanding of the dynamics of a magnetic skyrmion \cite{Muhlbauer09,Rossler06}, a particle-like topologically stable spin configuration, and more importantly its response dynamics to external fields and under resonant excitation are important subjects for the manipulation of skyrmions, relevant for practical applications \cite{Fert13,Romming13}. Spin-polarized currents have proven to be an efficient way to induce current-driven translational motion of skyrmions in metallic magnets \cite{Jonietz10,Iwasaki13,Lin13}. However, the observation of skyrmion phases in the multiferroic insulator Cu$_2$OSeO$_3$ \cite{Seki12,White12} necessitates the search for new means to manipulate the skyrmions in the absence of spin transfer torque. Among these methods are electric \cite{Liu13} and magnetic \cite{Wang17,Everschor12,Komineas15} field gradients, temperature gradients \cite{Kong13,Lin14,Mochizuki14}, magnon currents \cite{Iwasaki14,Schutte14,Zhang15,Zhang17}, and microwave fields \cite{Wang15,Moon16}. 

In addition, for the efficient manipulation of spin structures at the nanoscale it is necessary to have a better understanding of the dissipation processes present in the system. Classically, the dynamics of any magnetic texture is described by the Landau-Lifshitz-Gilbert (LLG) equation \cite{lifshitzBK80,gilbertTM04} which reduces to a particle-like equation of motion for the skyrmion collective coordinate known as Thiele's equation \cite{Thiele}. Dissipation mechanisms are included by a phenomenological local (in time) velocity-dependent term which parameterizes the coupling of the skyrmion to other degrees of freedom, such as electrons, magnons, or phonons. Quite recently these classical micromagnetic equations of motion were generalized for the quantum propagation of a skyrmion in a magnetic insulator \cite{Psaroudaki17}, where it has been shown that the interaction of the skyrmion with the magnon modes, the only low-energy relevant excitations in an insulating system, gives rise to dissipation described by a nonlocal damping kernel with memory effects. The derived equation of motion relies on the typical approach according to which the intrinsic spin degrees of freedom are divided into the system (skyrmion) and the surrounding environmental bath (magnetic excitations) governed by a dissipative process \cite{WeissBook}. The application of a time-dependent field is usually described as an external force which couples directly to the system, whereas the \textit{impact of the driving field on the environment} is usually neglected. However in many physical situations the driving force inevitably interacts with the environmental bath and is well known to result in important contributions to the dynamical response of nanoscale systems, for example in the case of driven superconducting tunnel junctions \cite{Grabert15,Frey16}, quantum two-level systems \cite{Reichert16,Grabert16} and quantum ratchets \cite{Denisov07,Kohler05}. 

In this Letter, we study the dynamics of a skyrmion in a magnetic insulating nanowire in the presence of a time-dependent oscillating field gradient which acts as an external driving force on the skyrmion. At the same time, the linear coupling of the external field with the magnetic excitations gives rise to time-dependent dissipation terms for the skyrmion's collective coordinate of position. Under resonance conditions the skyrmion undergoes a unidirectional helical propagation to the otherwise periodic bounded motion. The present formulation reveals that the time-dependent driving of the modes of the magnon bath results in a net force acting on the skyrmion which cannot be reproduced by the usual phenomenological theories. 

\begin{figure}[b]
\includegraphics[width=1\linewidth]{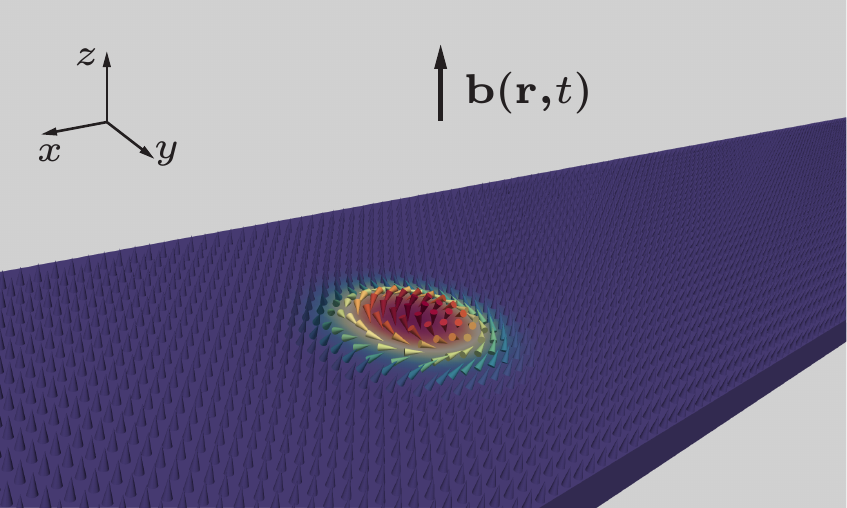}
\caption{A magnetic skyrmion with topological number $Q=-1$ placed in a nanowire strip in the $xy$-plane with a time-oscillating magnetic field $\mb{b}(\mb{r},t)$ applied along the perpendicular $z$-direction.}
\label{fig:Fig1}
\end{figure}
\begin{table*}
\caption{\label{Units} Physical Units for  $J=1$ meV, $J/D = 4$, $S=1$, and $\alpha= 5$ \AA.}
\begin{ruledtabular}
\begin{tabular}{ccccccccc}
 Length&Time&Frequency&Magnetic field&Force$\times N_A$&Spring constant$\times N_A$&Mass$\times N_A$&Friction$\times N_A$
\\ \hline
\\
 $ 2$ nm&$6.58\times 10^{-13}$ s &$241.8$ GHz &$0.54$ T&$8 \times10^{-14}$ N&$4\times 10^{-5}$ N/m&$ 0.17\times 10^{-28}$ kg& $5.27 \times 10^{-26}$ N s/m \\
\end{tabular}
\end{ruledtabular}
\end{table*}
\emph{Model}. The Euclidean action of a magnetic insulator in the two-dimensional (2D) $xy$ plane with a normalized magnetization 
$\mb{m} (\mb{r})=[\sin\Theta(\mb{r})\cos\Phi(\mb{r}),\sin\Theta(\mb{r})\sin\Phi(\mb{r}), \cos\Theta(\mb{r})]$ is given by
\begin{align}
S_{E} = N_A \int d\mb{r} \int dt \left(\bar{S}  \dot{\Phi} (1-\Pi) - \mathcal F(\Phi,\Pi) \right)\,,
\label{EucAction}
\end{align}
where $\bar{S}= S (J/D)^2$ with $S$ the magnitude of the spin, and $J$ and $D$ are the strength constants of the Heisenberg and the Dzyaloshinskii-Moriya (DM) interactions, respectively, measured in units of energy. The stable skyrmion configurations are attributed to the DM interaction which results in chiral couplings \cite{Bogdanov94}. Here $N_A$ counts the number of magnetic layers in the $z$ axis along which the magnetization is uniform. Moreover, $\dot\Phi$ is the time derivative of $\Phi(\mb{r},t)$ and $\Pi(\mb{r},t)\equiv\cos\Theta$ is canonically conjugate to $\Phi$. The translationally invariant energy density functional in reduced units is of the form
\begin{align}
\mc{F}(\mb{m})= \sum_{i=x,y} \left( \frac{\pt \mb{m}}{\pt r_i}\right)^2 + \mb{m} \cdot \nabla \times \mb{m}   - h m_z \,,
\label{FreeEnergy}
\end{align}
where $h$ is the strength of the uniform magnetic field applied along the $z$ direction. The relation between physical and dimensionless parameters is given in Table~\ref{Units} for a specific choice of $J=1$ meV, $J/D = 4$ and $S=1$, and lattice constant $\alpha= 5$ \AA. The functional \eqref{FreeEnergy} supports a skyrmionic static solution $\mb{m}_0(\mb{r})$, which is parametrized in polar coordinates $\mb{r}= (\rho \cos \phi, \rho \sin \phi)$ by the rotationally symmetric solutions $\Theta_0(\rho,\phi)=\Theta_0(\rho)$ with boundary conditions $\Theta_0(0)=\pi$ and $\Theta_0(\infty)=0$, and $\Phi_0(\rho,\phi)= \phi + \pi/2 $ \cite{Bogdanov94}. We use the following approximate function for the magnetization profile $\Theta_0(\rho)$:
\begin{align}
\Theta_0(\rho) = A \cos^{-1}\left[\tanh (\frac{\rho - \lambda}{\Delta_0}) \right] \,,
\label{Profile_Lrg1}
\end{align} 
where $A=\pi/\cos^{-1}(\tanh(-\lambda/\Delta_0))$ provided that the parameters $\lambda$, which is the skyrmion radius, and $\Delta_0$ are calculated numerically from the Euler-Lagrange equation of the stationary skyrmion, $\delta \mc{F}/\delta \Phi_0=0=\delta \mc{F}/\delta \Pi_0$. Note that magnetic skyrmions are characterized by a finite topological charge 
\begin{align}
Q_0=\frac{1}{4 \pi} \int d \mb{r}~ \mb{m} \cdot (\pt_x \mb{m} \times \pt_y \mb{m}) \,,
\end{align}
which describes a nontrivial mapping from the 2D magnetic system into the 3D spin space \cite{Wilczek83}. For the skyrmions considered here we have $Q_0=-1$. An external time-dependent field $\mb{b}(\mb{r},t)= \Theta(t-t_0) h_z x \cos(\wext t) \hat{z}$ couples to the magnetization as $S_B = N_A \int_{\mb{r},t}\mb{b}(\mb{r},t) \cdot \mb{m}(\mb{r},t)$. Throughout this work it is assumed $S_0 \gg S_B$, where $S_0 = \int_{\mb{r},t} \mc{F}(\Phi_0,\Pi_0)$ is the skyrmion's configuration energy. $\Theta(t-t_0)$ is the Heaviside step function and signals the onset of the external field at time $t_0$. The gradient strength  in physical units is obtained as $H_z=h_z \times 270$ mT/nm. For details on definitions and notations see Ref.~\cite{Supp1}. 

The magnetic fluctuations around the skyrmionic background are found by performing expansions of the form $\Pi_0+\eta$ and $\Phi_0+\xi$ into the energy functional of Eq.~\eqref{FreeEnergy} up to second order in $\eta$ and $\xi$. They appear as solutions of the eigenvalue problem (EVP) $\mc{H} \Psi_{n} = \varepsilon_{n} \sigma_z \Psi_{n}$, where $\mc{H}$ is the Hermitian Hamiltonian $\mc{H}=\delta_{\chi^\dagger}\delta_\chi\mathcal F |_{\chi=\chi^\dagger=0}$ and $\chi= 1/2 \binom{C_0 ~~i/ C_0}{C_0 ~-i/C_0} \binom{\eta}{\xi}$ with $C_0=\sin \Theta_0$. Solutions of the EVP include propagating scattering states with eigenfrequencies above the magnon gap $\ems=h$ induced by the magnetic field, as well as massive internal modes that are found for energies $0 < \varepsilon_n \leq \ems$ and correspond to deformations of the skyrmion into polygons \cite{SchuttePRB14}. For a detailed analysis of the magnon spectrum see Ref.~\cite{Supp1}. 
\\ 
\emph{Effective dissipation.} In order to describe the magnetization dynamics we employ a functional integral formulation in which the partition function is given by $Z=\int \mc{D} \Pi \mc{D} \Phi ~e^{-S_E}$, provided that $S_E$ is given by Eq.\eqref{EucAction}. Our task is to develop a formalism that provides a natural interpretation of a particle for the skyrmion described by a dynamical variable in contact with the bath of magnetic excitations. We thus divide the magnetization field into the classical part and a fluctuating part,
\begin{align} 
\Pi(\mb{r},t)&=\Pi_0(\mb{r}-\mb{R}(t))+\eta(\mb{r}-\mb{R}(t),t)\,, \nonumber \\
\Phi(\mb{r},t)&=\Phi_0(\mb{r}-\mb{R}(t))+\xi(\mb{r}-\mb{R}(t),t)\,,
\label{Transformation}
\end{align}
where $\mb{R}(t)$ corresponds to the collective coordinates of the position of the skyrmion. A finite perturbation theory up to second order in terms of $\eta$ and $\xi$ is constructed using a path-integral version of the Keldysh technique to deal with real-time dynamics \cite{KeldyshRev} and also paying particular attention to define the correct measure of path integration by employing the Faddeev-Popov techniques for collective coordinates \cite{SakitaBook,Braun96,Psaroudaki17}. When the bath degrees of freedom are integrated out we arrive at an effective action which upon minimization leads to an equation of motion for the classical path $\mb{R}(t)$.  For a skyrmion placed in a magnetic nanowire driven by a time-oscillating magnetic field gradient (see Fig.~\ref{fig:Fig1} for a visualization of the proposed setup), the equation of motion reads
\begin{align}
&\tilde{Q}_0 \epsilon_{ij} \dot{R}^{j}(t) + \int_{t_0}^{t} dt' ~\dot{R}^j (t') \gamma_{ji}(t,t')= \Fext^i(t) + \mc{K}_{ij} R^j(t) \,,
\label{EquationTime}
\end{align}
where $\epsilon_{ij}$ is the Levi-Civita tensor and summation over repeated indices $i,j=x,y$ is implied. 
\begin{figure}[t]\centering
\includegraphics[width=1\linewidth]{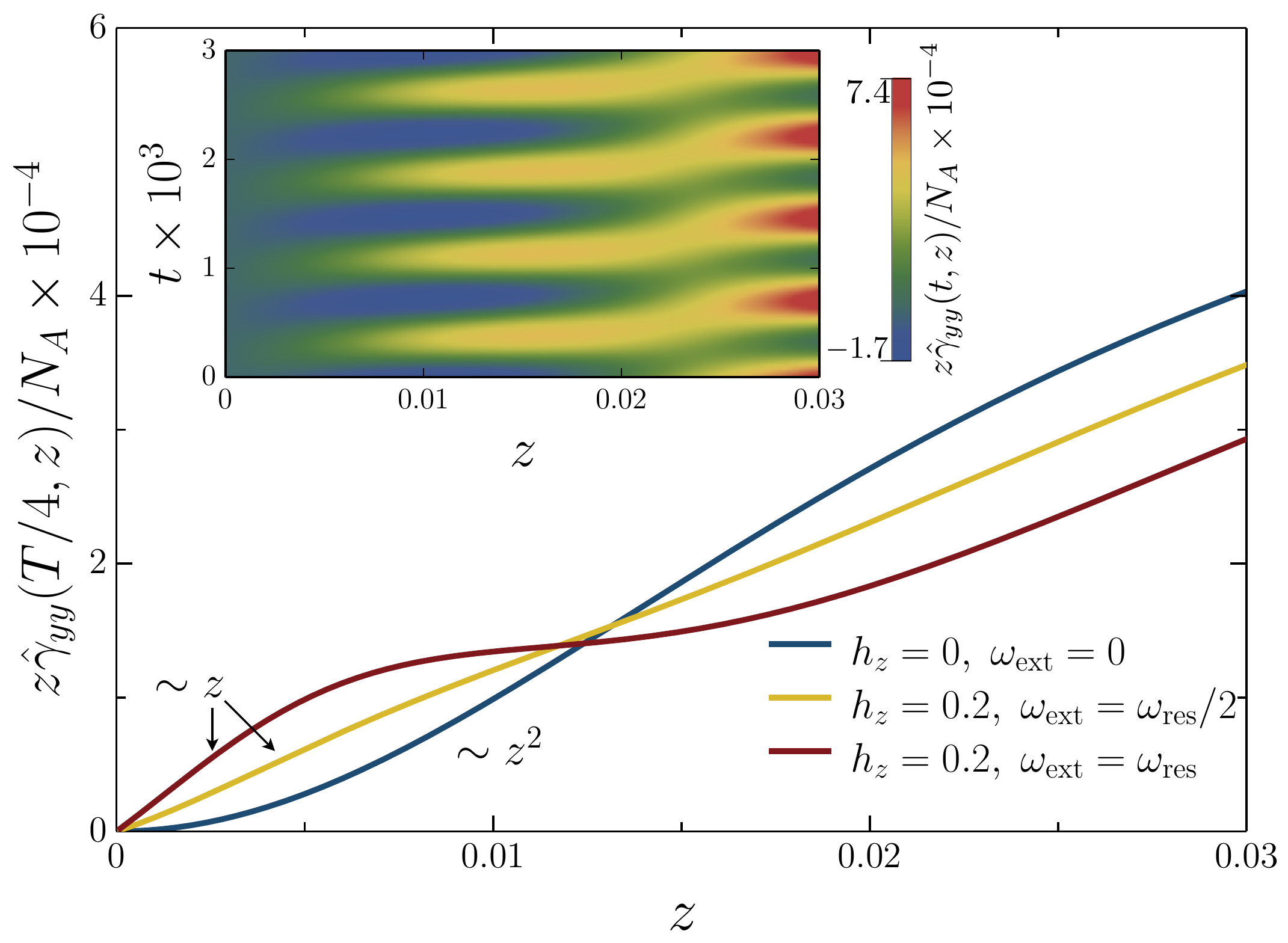}
\caption{ Frequency dependence of the diagonal part of the damping kernel $z \hat{\gamma}_{yy}(T/4,z)/N_A$ with $T=2 \pi /\wext$ and various values of $\wext$. The super-Ohmic to Ohmic transition manifests itself as different power law exponents of kernels $\hat{\gamma}_{yy}(t,z)$ and $\hat{\gamma}^0_{yy}(z)$ for finite external $\wext$. Inset: The colored surface  represents the quantity $z \hat{\gamma}_{yy}(t,z)/N_A$ for times $t$ up to $4 T$ and for $\wext = \wres/2$. Here $\wres=\varepsilon_0/\bar{S}$, with $\varepsilon_0=0.1358$ being the energy of the breathing mode and $\bar{S}=S (J/D)^2$ with $S=1$ and $J/D=4$.}
\label{fig:gyy}
\end{figure}
The first term in \eqref{EquationTime} is a Magnus force acting on the skyrmion proportional to the winding number $\tilde{Q}_0=4 \pi \bar{S} N_A Q_0$ ~\cite{Thiele,Stone96}. The nonlocal-in-time damping kernel $\gamma_{ji}(t,t')$ describes effective dissipation that originates from the coupling of the skyrmion to the magnon modes and is derived microscopically in a closed form
\begin{align}
\gamma_{ji}(t,t')= - N_A \bar{S}^2 \pt_{t'} \left( G^{A}_{ji}(t',t)+ G^{R}_{ij}(t',t) \right) \,,
\end{align}
where $G_{ij}^{A,R}(t,t') = \int_{\mb{r},\mb{r'}} f_i(\mb{r}) G^{A,R}(\mb{r},\mb{r}',t,t') f_{j}(\mb{r}')$ with the retarded and advanced propagator given by $G^{R,A}(\mb{r},\mb{r}',t,t')=[i \bar{S} \sigma_z \partial_t \pm i0 - \mc{H} - V(\mb{r},t)]^{-1}$. Here we define the functions $f_{i}= 1/2 \binom{s_i}{s_i^{\ast}}$ with $s_i=\sin \Theta_0 \pt_i \Phi_0 -i \pt_i \Theta_0$. The magnon Hamiltonian acquires an additional potential term $\Vconf(\mb{r})$ due to the confining potential, $\mc{H} \rightarrow \mc{H} + \Vconf$, while due to the time-dependent field the magnetic excitations experience the potential $V(\mb{r},t) =\mb{b}(\mb{r},t) \cdot \mb{D}$ with $ \mb{D} =  \delta_{\chi^\dagger}\delta_\chi \mb{m} |_{\chi=\chi^\dagger=0}$. The potential $V(\mb{r},t)$ is treated perturbatively within first order perturbation theory as $G^{R,A}(t,t') = G_0^{R,A}(t,t') + G_0^{R,A}  \circ V \circ G_0^{R,A}$, where we introduced the compact notation $ G \circ V \circ G \equiv \int_{\mb{r}_1, t_1} G(\mb{r},\mb{r}_1,t,t_1) V(\mb{r}_1,t_1)G(\mb{r}_1,\mb{r}',t_1,t') $. Note that generally the unperturbed propagators depend on relative times $G^{R,A}_0(\mb{r},\mb{r}',t,t') \equiv G^{R,A}_0(\mb{r},\mb{r}',t-t')$, a property which is not fulfilled by the full propagators $G^{R,A}$ due to the explicit time dependence of the potential $V(\mb{r},t)$. With this preparation it now becomes apparent that the damping kernel can be decomposed as follows 
\begin{align}
\gamma_{ji}(t,t') =\gamma_{ji}^{0}(t-t') +\Delta \gamma_{ji}(t,t-t') \,,
\end{align} 
where $\gamma_{ji}^{0}(t-t')$ contains contributions from the unperturbed Hamiltonian and $\Delta \gamma_{ji}(t,t-t')$ contains contributions from the oscillating field. A more transparent form of Eq.~\eqref{EquationTime} is obtained in Fourier space,
\begin{align}
(- i \omega) (\tilde{Q}_0 \varepsilon_{ij}+ \gamma_{ji}(t,\omega)) R^{j}(\omega) = \Fext^i(\omega) + \mc{K}_{ij} R^{j}(\omega)  \,,
\label{EquationFrequency}
\end{align}
provided that a time dependent function $g(t)$ is expanded in a Fourier series as $g(t) = \int_{-\infty}^{\infty} \frac{d\omega}{2 \pi} g(\omega) e^{-i \omega t}$ and that $\gamma_{ji}(t,\omega) = \gamma_{ji}^0(\omega) + \Delta \gamma_{ji}(t,\omega)$, with $\Delta \gamma_{ji}(t,\omega)$ being a $T$-periodic function of time with period $T=2\pi/\wext$. Explicit expressions of the damping kernels are evaluated on the basis of eigenfunctions of $\mc{H}$ and are presented in Ref.~\cite{Supp1}. From 
the full magnon spectrum we focus on four localized states found variationally with eigenfrequencies $\varepsilon_2 =0.0704 $, $\varepsilon_0 = 0.1358$, $\varepsilon_3 = 0.3358$, and $\varepsilon_{-1}=0.4498$. Higher in energy scattering states are neglected under the assumption that the effective dissipation depends on the inverse of the magnon energy $\varepsilon_n$ \cite{Psaroudaki17} and thus bound states have a larger contribution. 
\begin{figure}[t]\centering
\includegraphics[width=1\linewidth]{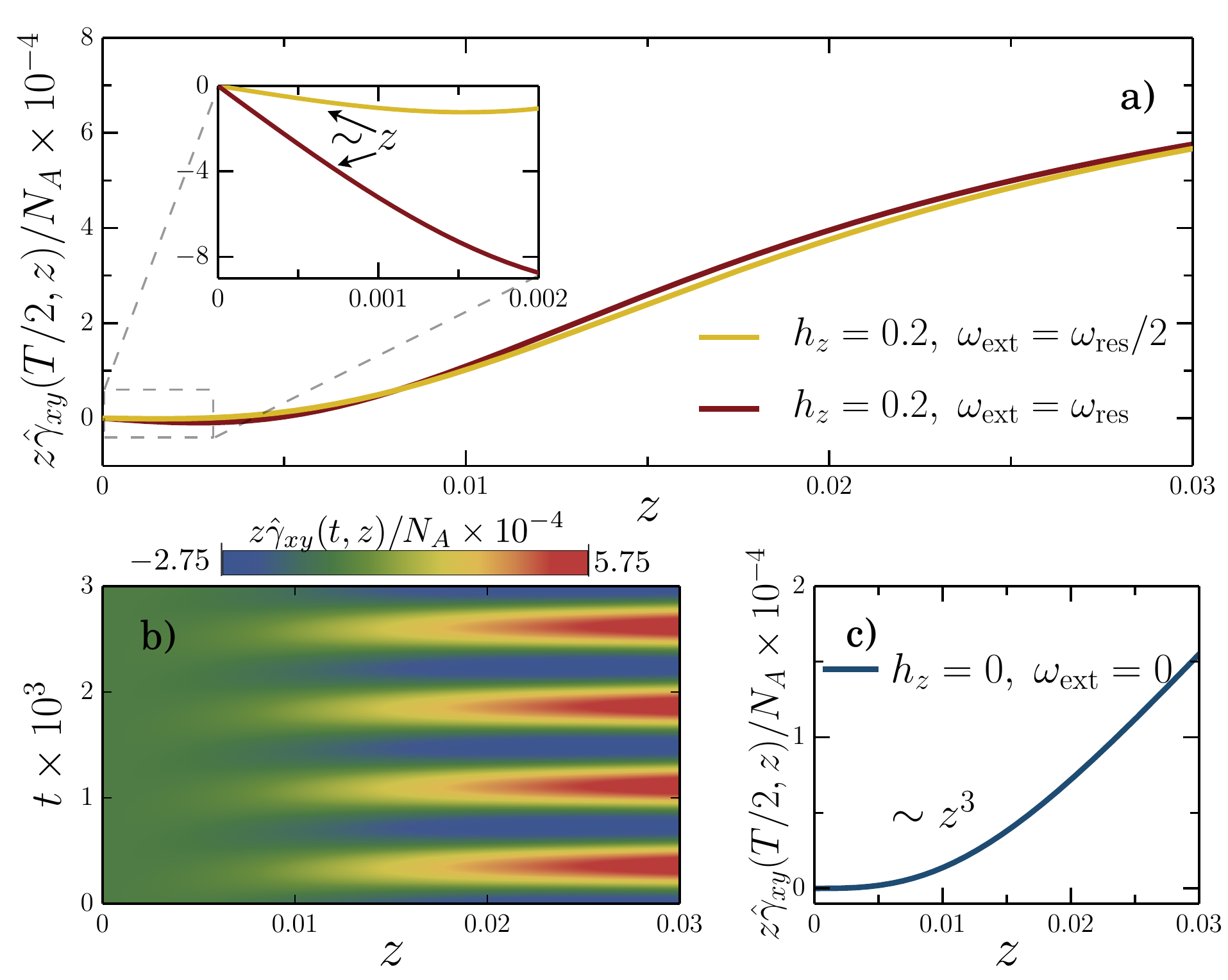}
\caption{a) Frequency dependence of the off-diagonal part of the damping kernel $z\hat{\gamma}_{xy}(T/2,z)/N_A$ with $T=2 \pi /\wext$ and various values of $\wext$. The super-Ohmic to Ohmic transition manifests itself as different power law exponents of kernels $\hat{\gamma}_{yy}(t,z)$ and $\hat{\gamma}^0_{yy}(z)$ depicted in c). In b) the colored surface represents the quantity $z \hat{\gamma}_{xy}(t,z)/N_A$ for times $t$ up to $4 T$ and for $\wext = \wres /2$. Here $\wres=\varepsilon_0/\bar{S}$, with $\varepsilon_0=0.1358$ the energy of the breathing mode and $\bar{S}=S J/D$ with $S=1$ and $J/D=4$. }
\label{fig:gxy}
\end{figure}

In Figs.~\ref{fig:gyy}-\ref{fig:gxy} we present the Laplace frequency dependence of the quantity $z \hat{\gamma}_{ij}(t,z)$, where we define $\hat{\gamma}_{ij}(t,z)= \gamma_{ij}(t,\omega = i z)$, at specific times $t$ and for different values of the external frequency $\wext$. Note that  $\hat{\gamma}_{yx}(t,z)=-\hat{\gamma}_{xy}(t,z)$ and that $\hat{\gamma}_{xx}(t,z)$ has a similar structure to $\hat{\gamma}_{yy}(t,z)$ and its analysis is omitted for brevity. In the absence of the external drive and at low frequencies $z \ll \egap/\bar{S}$, where $\egap$ is the lowest magnon energy, the diagonal kernel scales as $\gamma^0_{ii}(z) = \mc{M}_{ii} z+ \mc{O}(z^2)$, thus the main effect of the environmental coupling on the asymptotic dynamics is an induction of an effective mass $\mc{M}_{ii}$ predicted in \cite{Psaroudaki17}. On the contrary, the off-diagonal kernel has a subleading behavior $\gamma^0_{xy}(z) \sim z^2$, irrelevant for the asymptotic skyrmion  dynamics (see Fig.~\ref{fig:gxy}-c). It is now evident that the most unexpected feature of Figs.~\ref{fig:gyy}-\ref{fig:gxy} is a \textit{super-Ohmic to Ohmic} crossover behavior signaled by a low-frequency \textit{linear} dependence of the quantity $z \hat{\gamma}_{ij}(t,z)$ which corresponds to the Ohmic friction proportional to the skyrmion velocity. These Ohmic components are time-dependent and are formally defined below. The periodic behavior of the damping kernels with respect to time is visualized in Figs.~\ref{fig:gyy}-\ref{fig:gxy} where we plot the quantity $z \hat{\gamma}_{ij}(t,z)$ for times up to $t=4 T$ and for $\wext= \wres/2$ (colored surfaces).  
\begin{figure}[t]\centering
\includegraphics[width=1\linewidth]{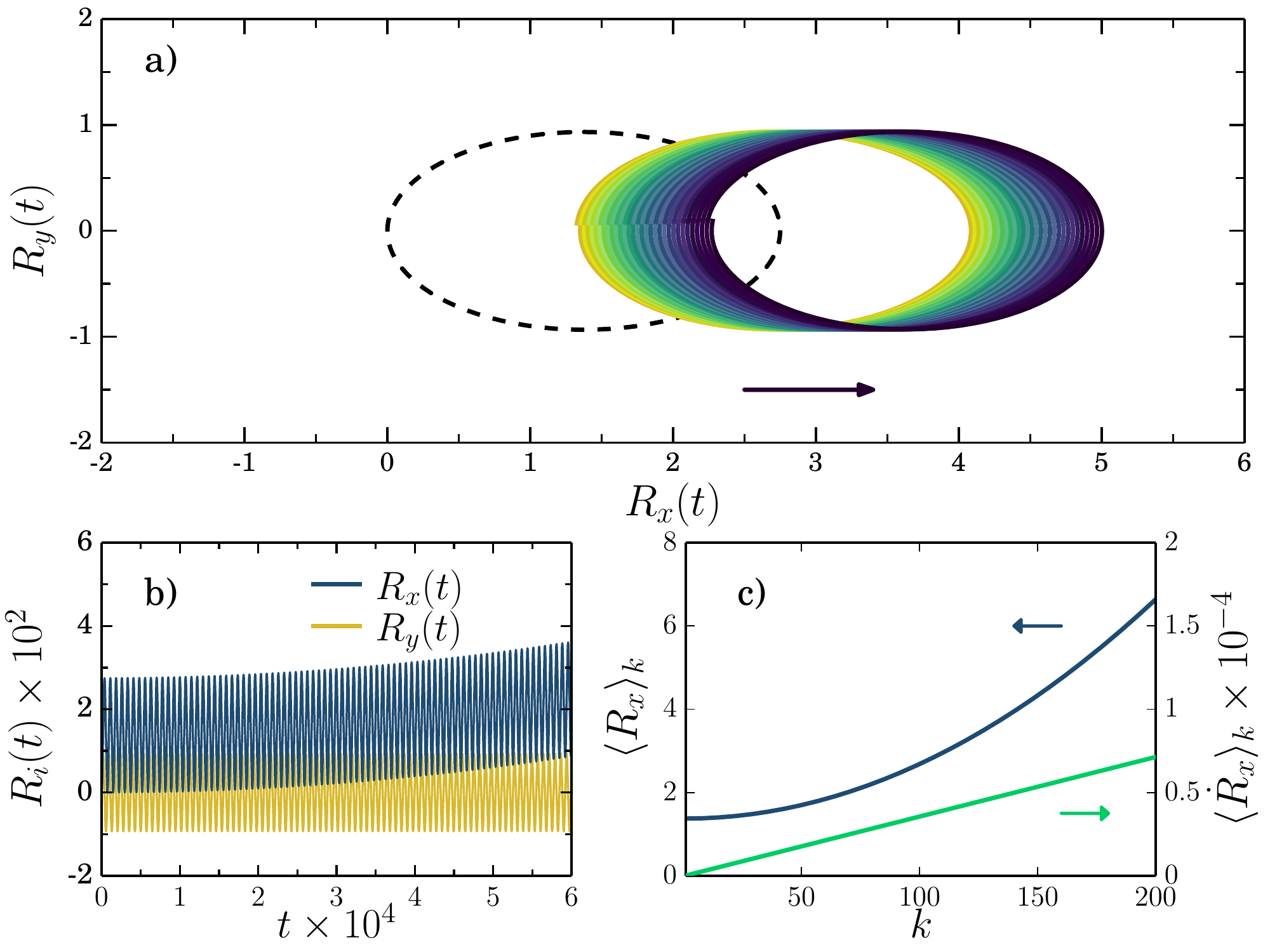}
\caption{a) Helical path for the skyrmion's collective coordinate $R_{i}(t)$ calculated by a numerical solution of the equation of motion \eqref{EquationReduced}. The inclusion of the time-dependent dissipation induces a unidirectional motion to the otherwise $T$- periodic bounded motion (black dashed line), where $T=2\pi/\wext$. The displacement of the skyrmion can be visualized by the sequential coloring starting from $t=100 T$ (yellow line) up to $t=130 T$ (black line). The black arrow shows the direction of motion. b) $T$-periodic time dependence of the collective coordinates $R_i(t)$ and c) mean position $\langle R_x \rangle_k$ and velocity $\langle \dot{R}_x \rangle_k$ as a function of consecutive periods $k$. }
\label{fig:Path}
\end{figure}

For the rest of the terms appearing in Eq.~\eqref{EquationTime}, the external time-dependent force equals $\Fext^i(t)/N_A = \frac{\delta}{\delta R_i} \int d\mb{r} ~\mb{b}(\mb{r},t) \cdot \mb{m}_0(\mb{r}-\mb{R})$, whereas the spring constant $\mc{K}_{ij}= \delta_{ij} K^j$ parametrizes the effects of the lateral parabolic confining potential $\Wconf(\mb{r})$ of a linear track and is defined as $K^i/N_A = \frac{\delta^2}{\delta R_i^2} \int d \mb{r} \Wconf(y-R_i) $. Throughout this work we use $h=0.5$ ($270$ mT), and for this choice we find that the skyrmion size is $\lambda = 3.3$ ($6.6$ nm) and $\Delta_0=1.2$ ($2.4$ nm). In addition, we choose $\mb{b}(\mb{r},t) = \Theta(t-t_0) h_z x \cos(\wext t) \hat{z}$ which yields a linear force $\Fext^x(t)/N_A=\Theta(t-t_0) h_z F_0 \cos(\wext t) $ with $F_0=-79.71$, while $\Fext^y(t)=0$. The onset of the external field at time $t_0$ coincides with the time of preparation of the initial state and without loss of generality we set $t_0 = 0$. The geometry considered is the one of a narrow strip of width $4 \lambda$ in the $y$ direction and infinite in the $x$ direction. In such a linear track the repulsive forces imposed by surface twists create a potential of the form $\Wconf(y) = 2 \cosh( y h/2)$ \cite{Meynell14}. For $h=0.5$ we find $K^y/N_A = 2.51$ and $K^x =0$. 

Note that Eq.~\eqref{EquationTime} contains only temperature-independent semiclassical terms proportional to the effective spin $N_A S$, while the temperature-dependent quantum terms $\propto \mc{O}[(N_A S)^0]$ are neglected under the assumption $N_A S \gg 1$. We also omit random forces $\zeta_{i}(t)$ that naturally appear due to the coupling of the skyrmion with the bath of magnons at finite temperature with the vanishing classical ensemble average $\langle \zeta_i(t) \rangle_{\scalebox{0.7}{cl}}=0$ \cite{WeissBook}. We thus implicitly assume $\mb{R}(t)= \langle \mb{R}(t) \rangle_{\scalebox{0.7}{cl}}$. The Langevin equation for the skyrmion, including temperature-dependent quantum dissipation under an external oscillating field is presented in a forthcoming work .  
\\
\emph{Helical motion.}
To study the skrymion propagation under resonant excitation we consider the case when the external frequency $\wext$ is close to the eigenfrequency of the breathing mode $\wres=\varepsilon_{0}/\bar{S}$, which is the most relevant mode under the application of out-of-plane ac magnetic field \cite{Mochizuki12,Onose12,Okamura13,Kim14}. Resonance conditions require the inclusion of a relaxation rate $\Gamma$, {\it i.e.}  $\varepsilon_{n} \rightarrow \varepsilon_n - i \Gamma$, which phenomenologically parametrizes any mechanism that could lead to relaxation of the magnon bath dynamics, for example the back-action of the skyrmion to the bath, magnon-magnon and phonon-magnon interaction.  
As long as we are interested in asymptotic times $t-t' \gg \bar{S} \Gamma^{-1}$ and $t-t' \gg \bar{S} \egap^{-1}$; therefore for $\omega \ll \Gamma/\bar{S}$ and $\omega \ll \egap/\bar{S}$, we can focus on the low frequency power law behavior of the friction kernel $\gamma_{ij}(t,\omega)$, {\it i.e.} we perform a Taylor expansion around the origin as $\gamma_{ij}(t,\omega) \simeq \gamma_{ij}(t,0) +(- i \omega) \pt_{\omega} \gamma_{ij}(t,\omega)\vert_{\omega =0}  +\mc{O}(\omega^2)$. Upon substituting this low-frequency behavior of dissipation kernels in Eq.~\eqref{EquationFrequency} and performing an inverse Fourier transform we arrive at a local equation of motion
\begin{align}
 \begin{bmatrix}
D_x(t) + M_x(t) \pt_t& Q(t) + G(t) \pt_t\\
-Q(t) -G(t)\pt_t & D_y(t) + M_y(t) \pt_t
\end{bmatrix} \binom{\dot{R}_x}{\dot{R}_y}  = \binom{\Fext^x(t)}{K^y R_y} \,,
\label{EquationReduced}
\end{align}
where $D_i (t) = \delta D_i \sin(\wext t)$, $G(t) = \delta G \sin(\wext t)$, $M_{i}(t) =  \mc{M}_{ii}+ \delta M_i \cos(\wext t)$ and $Q(t)= \tilde{Q}_0 +\delta Q \cos(\wext t) $. A detailed derivation of the above simplified dynamics as well as exact expressions in closed formulas for the components $\delta D_i$, $\delta G$, $\delta M_i$ and $\delta Q$ can be found in Ref.~\cite{Supp1}. In  Fig.~\ref{fig:Path} we present a numerical solution of Eq.~\eqref{EquationReduced} using $\wext= \varepsilon_0/\bar{S}= 8.49 \times 10^{-3}$ ($2.05$ GHz), $h_z= 0.02$ ($5.4$ mT nm$^{-1}$), and $\Gamma = \wext \times10^{-3}$. For these values we find $\mc{M}_{yy}/N_A= 1.17$, $\mc{M}_{xx}/N_A=0.14$, $\delta M_{y}/N_A = -0.72 $, $\delta M_{x}/N_A = -0.10 $, $\delta Q/N_A = -5.51 \times 10^{-4}$, $\delta D_x/N_A = 5.54 \times 10^{-4} $, $\delta D_y/N_A = 4.14  \times 10^{-3}$ and $\delta G = -0.22 $.  The physical units for these parameters can be found in Table~\ref{Units}. Note that  Eq. \eqref{EquationReduced} is solved with initial conditions $R_i(t_0) = 0$ and $\dot{R}_i(t_0)=0$. 

To identify the effect of the driving of the magnon bath, we first consider the skyrmion motion when all the time-dependent dissipation terms appearing in Eq.~\eqref{EquationReduced} vanish. In this case, the skyrmion performs a bounded periodic motion in the $x$ direction with period $T=2\pi/\wext=740$ around the equilibrium position $\Req \simeq -h_z F_0 K^y /(\tilde{Q}_0^2 \wext^2)=1.37$ (black dashed line). The behavior of Fig.~\ref{fig:Path}-b) implies that the time-dependent dissipation induces a unidirectional helical motion for the skyrmion along the long axis of the nanowire. To further explore this effect we define the average position as $\langle R_{i} \rangle_k =1/T \int_{ k T}^{(k+1) T} R_{i}(t) dt$ and we find that $\langle R_y \rangle_k \simeq 0$ along the short axis and $\langle R_{x} \rangle_k = 1.37+ 1.32 \times 10^{-4} k^2$. Therefore, as an approximate solution of Eq.~\eqref{EquationReduced} we use the ansatz $\langle R_{x} \rangle_k \simeq \Req + k^2 C_1$, with $C_1= -h_z F_0 T^2(\wext \delta G + K^y \delta M_x/\tilde{Q}_0 -\delta Q)/(2 \tilde{Q}_0 ^3) $. The mean position $\langle R_{x} \rangle_k $ is presented in Fig.~\ref{fig:Path}-c) as a function of consecutive periods $k$. For the mean velocity we find $\langle \dot{R}_x \rangle_k = 3.57 \times 10^{-7} k $ and $\langle \dot{R}_y \rangle_k \simeq 0$. To give an estimate in physical units, for $k=150$ and thus within $73$ ns the skyrmion is displaced by $5.94$ nm from its initial equilibrium position with mean velocity $16.3$ cm/s. Note that the direction of motion as well as the sign of the mean velocity is independent of the sign of the external ac field $h_z$. Any sense of direction is defined by the sign of the topological charge $\tilde{Q}_0$; we observe a displacement along the $\pm x$ axis with positive/negative mean velocity for a skyrmion with $\mp \vert \tilde{Q}_0	 \vert$ charge. \\
\emph{Conclusion.} We have shown that the skyrmion motion is helical under the driving of an oscillating magnetic field of high symmetry, see Fig.~\ref{fig:Path}. Remarkably, this unidirectional motion is generated by the action of the field on the magnon bath 
formed by the magnetic excitations of the skyrmion. 
This stands in contrast to setups in which the
axial symmetry is explicitly broken by magnetic fields with in-plane components. In this case, a unidirectional motion directly follows from the phenomenological LLG \cite{Wang15,Moon16}.
Here, however, the driving of the magnons gives rise to a time-dependent dissipation kernel,  resulting in a helical skyrmion dynamics that 
 cannot be obtained from the usual theories with a phenomenological friction term. 
 Our predictions can be tested experimentally in magnetic nanowires.

\begin{acknowledgments}
\emph{Acknowledgments.} We are grateful to Pavel Aseev and Sebastian Diaz for useful discussions. This work was supported by the Swiss National Science Foundation and NCCR QSIT.
\end{acknowledgments}

\pagebreak
\widetext
\begin{center}
\textbf{\large Supplemental Material for\\Skyrmions Driven by Intrinsic Magnons}
\end{center}

\section{A. Model } 
\sloppy We consider a thin magnetic insulator with normalized magnetization $\mb{m}(\tr,\tilde{t} )=[\sin \Theta (\tr,\tilde{t} ) \cos \Phi (\tr,\tilde{t} ),\sin \Theta (\tr,\tilde{t} ) \sin \Phi (\tr,\tilde{t} ), \cos \Theta (\tr,\tilde{t} ) ]$ which is described by the real-time action  
\begin{align}
S= \frac{S N_A}{\alpha^2} \int d\tilde{t} \int d \tr ~\dot{\Phi} (\Pi-1) - N_A \int d\tilde{t}  \int d\tr \mc{W}(\Phi,\Pi) ,
\end{align}
where $\tr = (\tilde{x},\tilde{y})$, $S$ is the magnitude of the spin, $N_A$ is the number of magnetic layers along the $z$ axis and $\alpha$ is the lattice spacing. The real-time derivative of field $\Phi$ is denoted as $\dot{\Phi}$, while the field $\Pi \equiv \cos \Theta$ is canonically conjugate to $\Phi$. The skyrmion configurations are metastable solutions of the following energy functional 
\begin{align}
\mc{W}(\mb{m})= \sum_{i=x,y} J \left( \frac{\pt \mb{m}}{\pt \tilde{r}_i}\right)^2 +\frac{D}{\alpha} \mb{m} \cdot \nabla_{\tilde{\mb{r}}} \times \mb{m} - H m_z \,,
\label{FreeEnergy}
\end{align}
where $J$ and $D$ have units of energy and $H= g\mu_{B} \tilde{H}/\alpha^2$, where $\tilde{H}$ is in units of T. The energy functional is transnationally invariant, thus the model has two zero energy modes $\mc{Y}_i$ associated with translations along the $i=\tilde{x}$ and $i=\tilde{y}$ directions. The application of a time-dependent nonuniform magnetic field $\mb{B}(\tr,\tilde{t})= \Theta(\tilde{t}-\tilde{t}_0) \Hext \tilde{x} \cos(\wextt \tilde{t}) \hat{z}$ gives additional terms in the energy functional of the form $S_B= \int d\tilde{t}  \int d\tr~ \mb{B}(\tr,\tilde{t}) \cdot \mb{m}(\tr,\tilde{t})$, which upon minimization will generate external forces acting on the skyrmion configuration. Here it holds that $\Hext = g \mu_B \tilde{H}_{\mbox{\scriptsize{ext}}}/\alpha^3$, with $\tilde{H}_{\mbox{\scriptsize{ext}}}$ in T. 

We introduce dimensionless variables $\mb{r}= \tr/(l\alpha)$, $t=\tilde{t} J S^2$ and $\beta=\tilde{\beta} J S^2=\frac{J S^2}{k_B T}$, where $T$ is the system temperature, $k_B$ is the Boltzmann constant and $l=J/D$. Also note that throughout this work we assume $\hbar=1$. We define the energy functional in reduced units as
\begin{align}
\mc{F}(\mb{m})= \frac{(l \alpha)^2}{J} \mc{W}(\mb{m}) = \sum_{i=x,y} \left( \frac{\pt \mb{m}}{\pt r_i}\right)^2 + \mb{m} \cdot \nabla_{\mb{r}} \times \mb{m} - h m_z
\end{align}
with $h=(l \alpha)^2 H/J= J g\mu_B \tilde{H}/D^2$. Similarly we find that $S_B= \int dt  \int d \mb{r}~ \mb{b}(\mb{r},t) \cdot \mb{m}(\mb{r},t)$, where $\mb{b}(\mb{r},t)= \Theta(t-t_0) h_z x \cos(\wext t) \hat{z}$ with $h_z = (l \alpha)^3 \Hext /J$. Thus the strength of the field gradient is $H_z=\tilde{H}_{\mbox{\scriptsize{ext}}}/\alpha= (D^3 h_z)/(\alpha g \mu_B J^2)$.  

Stationary configurations of the action, denoted as $\Phi_0(\mb{r})$ and $\Pi_0(\mb{r})$, are found by minimising the energy functional, \textit{i.e.} by solving the coupled equations $\frac{\delta \mc{F}}{\delta \Phi_0}=0=\frac{\delta \mc{F}}{\delta \Pi_0}$. This class of solutions is characterized by a finite topological charge $Q$
\begin{align}
Q= \frac{1}{4 \pi} \int d\mb{r} ~\mb{m} \cdot (\pt_x \mb{m} \times \pt_y \mb{m}) ,
\label{Winding}
\end{align}
and a magnetization profile $\Theta_0$ that decays to zero at spatially infinity. Note that the static configuration is defined in the absence of the driving force. 

\section{B. Collective coordinates and Quantum fluctuations} 

The dynamics of the skyrmion is described by the collective coordinate fields $\mb{R}(t)$ and fluctuations around the static solutions
\begin{align}
\Phi(\mb{r},t)&= \Phi_0(\mb{r} -\mb{R}(t)) + \xi(\mb{r}-\mb{R}(t),t) \nonumber \\
\Pi(\mb{r},t)&= \Pi_0(\mb{r} -\mb{R}(t)) + \eta(\mb{r}-\mb{R}(t),t) \,. 
\label{Transformation}
\end{align}
For reasons of convenience we describe the fluctuations using the following spinor notation, 
\begin{equation}
\chi =\frac{1}{2} \binom{\xi \sin \Theta_0+i \eta/ \sin\Theta_0 }{ \xi \sin \Theta_0-i \eta/ \sin\Theta_0} \,.
\label{SpinorNotation}
\end{equation}
The action \eqref{Action} under transformation \eqref{Transformation} and in the language of spinors $\chi$ and $\chi^{\dagger}$ takes the form
\begin{align}
S=  N_A  \int dt \int d \mb{r} \left( -\bar{S} \dot{\mb{R}} \Pi_0 \nabla \Phi_0 + \chid\cdot (\mc{G}+\mc{K}) \chi + \mc{J}^{\dagger} \cdot \chi + \chid \cdot \mc{J} \right) ,
\label{Action}
\end{align}
 where $\bar{S}=S l^2$ and we introduce the compact scalar product notation for operators and functions
\begin{align}
\chi^{\dagger}\cdot \mathcal{G}\chi\ &\equiv \int d\tau d\tau' d \mb{r}  d \mb{r}' \, \chi^{\dagger} (\mb{r},\tau)\mathcal{G}(\mb{r}, \tau; \mb{r'},\tau') \chi (\mb{r'},\tau')\,,
\end{align}
and the scalar product in spinor space is left implicit. Further we define $\mc{G}=  i \bar{S}\sigma_z \partial_{t} - \mathcal{H}$, $\mc{K} =   -i \bar{S} \sigma_z   \dot{R}_i \Gamma_i$, and $\mc{J}=- i\sigma_z \bar{S} \dot{R}_i f_i $.  Repeated indices, $i,j=x,y$, are summed over and we introduce the abbreviation $\Gamma_i= \pt_i  - \sigma_x \cot \Theta_0 \pt_i  \Theta_0$, while magnon Hamiltonian $\mc{H}$ is defined as $\mathcal H=\delta_{\chi^\dagger}\delta_\chi\mathcal F |_{\chi=\chi^\dagger=0}$. Functions $f_i$ are defined in the spinor language as $f_i=\frac{1}{2}\binom{s}{ s^*}$, with $s=\sin \Theta_0 \pt_i \Phi_0 - i \pt_i \Theta_0$. Note that we ignore an overall constant from the configuration energy $N_A \int dt \int d\mb{r} ~\mc{F}(\Phi_0,\Pi_0)$ and a gauge dependent boundary term proportional to $N_A \bar{S} \int dt \int d\mb{r} ~\dot{\mb{R}} \cdot \nabla \Phi_0$ that originates from the dynamical part of the action \eqref{Action}. The most common class of solutions of magnetic skyrmions has vanishing topological term with $\int d\mb{r}~ \nabla \Phi_0 = 0$. Similarly for the field gradient we get
\begin{align}
S_B = N_A \int dt \int d\mb{r}~ \left(\mb{b} (\mb{r},t) \cdot \mb{m}_0(\mb{r}-\mb{R}(t)) + \chid \cdot V(\mb{r},t) \chi + j^{\dagger} \chi + \chid j \right) \,,
\label{ExtAction}
\end{align} 
where $V(\mb{r},t) = \mb{b}(\mb{r},t) \cdot \mb{D}$ is the time dependent perturbation to the magnon Hamiltonian and $j=\mb{b}(\mb{r},t) \cdot\mb{j}$ corresponds to linear in fluctuations terms which are present because the static solution is defined in the absence of the time-dependent magnetic field. Also we define $ \mb{D} =  \delta_{\chi^\dagger}\delta_\chi \mb{m} |_{\chi=\chi^\dagger=0}$ and $\mb{j}= \delta_{\chi^\dagger} \mb{m} |_{\chi=\chi^\dagger=0}$. Here we assume that the skyrmion's configuration energy $S_0 = \int d\mb{r} \int dt \mc{F}(\Phi_0,\Pi_0)$ is much larger than the energy $S_B= \int d\mb{r} \int dt \mb{b}(\mb{r},t) \cdot \mb{m}(\mb{r},t)$ added by the external applied force, $S_0 \gg S_B$. In this case, the skyrmion configuration $\mb{m}_1(\mb{r},t)$, solution of the minimization of the action $S_0+S_B$, is $\mb{m}_1(\mb{r},t) \simeq \mb{m}_0(\mb{r}) $ at any $t$. Thus, the linear in fluctuation terms $j = \mb{b}(\mb{r},t)\cdot \delta_{\chi^{\dagger}} \mb{m} \vert_{\chi=\chi^{\dagger}=0}$ vanish. 

\section{C. The Keldysh functional integral}
We use the path-integral version of Keldysh technique \cite{KeldyshRev}, which is very useful when dealing with real-time dynamics at finite temperatures. The main idea is that integration over time is now replaced by integration over the Keldysh contour which consists of two branches. The upper branch goes from $t=-\infty$ to $t=+\infty$, and then the lower branch goes backwards from $t=+\infty$ to $t=-\infty$. It is convenient to split the bosonic field $\Phi$ into the two components $\Phi_+ \equiv \Phi(t+i0)$ and $\Phi_- \equiv \Phi(t-i0)$, that reside on the upper and the lower parts of the time contour, respectively. Similarly, we define fields $\Pi_{\pm}=\Pi(t \pm i 0)$. While the fields $\Phi$ and $\Pi$ are split into two components, it is convenient to assume that for the external perturbation it holds $\mb{b}(t+i0) = \mb{b}(t-i0) = \mb{b}(t)$.
The full form of fields $\Phi_{\pm}$ and $\Pi_{\pm}$ is 
\begin{align}
\Phi_{\pm}&= \Phi_0^{\pm}(\mb{r}-\mb{R}_{\pm}(t)) + \xi_{\pm}(\mb{r}-\mb{R}_{\pm}(t),t) \nonumber \\
\Pi_{\pm}&= \Pi_0^{\pm}(\mb{r}-\mb{R}_{\pm}(t)) + \eta_{\pm}(\mb{r}-\mb{R}_{\pm}(t),t) \,,
\end{align}
while the action \eqref{Action}-\eqref{ExtAction} becomes
\begin{align}
\mc{S}=S-S_B&= N_A \int dt d\mb{r}~\sum_{s=+,-} s\left( \bar{S} \dot{\Phi}_s (\Pi_s-1) - \mc{F}(\Phi_s, \Pi_s) -\mb{b} \cdot \mb{m} (\Phi_s, \Pi_s)  \right) \nonumber \\
&= N_A \sum_{s=+,-}s  \left( \int dt d\mb{r}~[-\bar{S} \dot{\mb{R}}_s \Pi_0^{s} \nabla \Phi_0^{s} -\mb{b} \cdot \mb{m}(\Phi_0^{s},\Pi_0^s) ] +\chi^{\dagger}_s \cdot (\mc{G}+ \mc{K}_s) \chi_s + \mc{J}^{\dagger}_s \cdot \chi_s + \chid_s \cdot \mc{J}_s \right) \,,
\label{ActionPM}
\end{align}
where the magnon Green's function is redefined as $\mc{G}= i \bar{S}\sigma_z \partial_{t} - \mathcal{H} -V(\mb{r},t)$. The partition function is given by
\begin{align}
Z= \int \prod_{s=+,-} \mc{D} \Phi_s \Pi_s e^{i\mc{S}} \,.
\end{align}
The gauxe fixing conditions in order to avoid overcounting degrees of freedom \cite{SakitaBook} are $G_i^s=\int d\mb{r} \chid_s \sigma_z \mc{Y}_i$, where $\mc{Y}_{i=x,y}$ are a degenerate pair of zero-frequency modes associated with the translational symmetry of the Hamiltonian $\mc{H}$. The gauxe-fixing conditions are implemented by inserting a factor of unity $\mathds{1} = \int \mc{D} \mb{R}_s \mbox{det}(J_{\mb{G}}) \delta(G^s_x)\delta(G_y^s)$ in the partition function 
\begin{align}
Z= \int \prod_{s=+,-} \mc{D} \chi_s \mc{D} \chid_s \mc{D} \mb{R}_s \det(J_{\mb{G}}^s) \delta(G_x^s)\delta(G_y^s) e^{i\mc{S}} \,. 
\end{align}
Here $J_{\mb{G}}^s (t,t')= d \mb{G}^s(t)/d\mb{R}(t')$ is the Jacobian of the coordinate transformation and is treated as an additional term in the action, adaptable for perturbative calculation. The next step is to perform a Keldysh rotation
\begin{align}
\chi_{c,q} = \frac{\chi_+ \pm \chi_-}{\sqrt{2}}\,, \qquad \mb{R}_{c,q} = \frac{\mb{R}_+ \pm \mb{R}_-}{\sqrt{2}} \,,
\label{Rotation}
\end{align}
where the subscripts $c,q$ denote the classical and the quantum components of the fields, respectively. The action \eqref{ActionPM} is of the form $\mc{S}= \mc{S}_{\mbox{\tiny{cl}}} + \mc{S}_{\mbox{\tiny{fl}}}$ where
\begin{align}
\mc{S}_{\mbox{\tiny{cl}}} =  N_A \int dt d\mb{r}~ \sum_{s=+,-} s\left(-\bar{S} \dot{\mb{R}}_s \Pi_0^{s} \nabla \Phi_0^{s} -\mb{b} \cdot \mb{m}(\Phi_0^{s},\Pi_0^s) \right) 
\label{ActionCl}
\end{align}
is the fluctuation-free part of the action, while 
\begin{align}
\mc{S}_{\mbox{\tiny{fl}}} = N_A  \left( \zeta^{\dagger} \cdot \hat{G} \zeta +\hat{J}^{\dagger} \cdot \zeta + \zeta^{\dagger} \cdot \hat{J}  \right)
\label{ActionFl}
\end{align}
is the fluctuation-dependent part. Field $\zeta = \binom{\chi_c}{ \chi_q}$ is introduced in order to obtain the action in a more compactified form. Further we define $\hat{G} =(\mc{G} +\frac{1}{\sqrt{2}} \mc{K}_c ) \sigma_x + \frac{1}{\sqrt{2}} \mc{K}_q \mathds{1}$ and $\hat{J}= \binom{\mc{J}_q}{\mc{J}_{c}}$. The partition function, neglecting terms $O(1)$ in $N_A$ which originate from the determinant $\mbox{det}(J_{\mb{G}}^s)$, is equal to
\begin{align}
Z=\int \mc{D} \mb{R}_c \mc{D} \mb{R}_q e^{i\mc{S}_{\mbox{\tiny{cl}}}}~[\mc{D} \zeta^{\dagger} \mc{D} \zeta \delta(G_x^c)\delta(G_y^c)\delta(G_x^q)\delta(G_y^q) e^{i \mc{S}_{\mbox{\tiny{fl}}}} ]  \,. 
\end{align}
The path integral over $\zeta$ and $\zeta^{\dagger}$ can be reduced to a Gaussian form by completing the square $\tilde{\zeta} = \zeta + \hat{G}^{-1} \hat{J}$. After integration, the partition function reduces to 
\begin{align}
Z=\int \mc{D} \mb{R}_c \mc{D} \mb{R}_q e^{i(\mc{S}_{\mbox{\tiny{cl}}} - N_A \hat{J}^{\dagger}\cdot \hat{G}^{-1} \hat{J})} \frac{1}{\det'({ N_A (-i \hat{G}}))} = \int \mc{D} \mb{R}_c \mc{D} \mb{R}_q e^{ i[\mc{S}_{\mbox{\tiny{cl}}} - N_A \hat{J}^{\dagger} \cdot \hat{G}^{-1} \hat{J} +i\mbox{\footnotesize{Tr}}'\mbox{\footnotesize{log}}(N_A (-i\hat{G})) ]} \,,
\label{PartitionFunc}
\end{align} 
where the prime on the determinant and the trace excludes the zero modes. The standard way to calculate the quasiclassical equation of motion for the skyrmion coordinate $\mb{R}_c$ is to calculate the saddle point of the action by extremizing with respect to the quantum coordinate $\mb{R}_q$. 

\section{C. Derivation of equation of motion for the classical path}

The minimization of action $\mc{S}_{\mbox{\tiny{cl}}}$ given in Eq.\eqref{ActionCl} with respect to quantum coordinate $\mb{R}_q$ results
\begin{align}
\frac{\delta \mc{S}_{\mbox{\tiny{cl}}}}{\delta R_q^j(t'')} = -4 \pi N_A Q \bar{S} \epsilon_{ij} \dot{R}_c^{j}(t'') -  N_A \int d\mb{r} ~\mb{b}(\mb{r},t) \cdot \partial_j \mb{m}_0 (\mb{r} - \mb{R}_c(t'')) \,,
\label{EquationCl}
\end{align}
where $Q$ is the skyrmion's topological number defined in Eq.\eqref{Winding} and $\epsilon_{ij}$ is the antisymmetric Levi-Civita tensor. Note that Eq.\eqref{EquationCl} is derived by expanding in quantum coordinate $\mb{R}_q$ up to first order. Next we calculate the equation of motion which results from the $\mc{S}_1=-N_A \hat{J}^{\dagger} \hat{G}^{-1} \hat{J}=- \hat{J}^{\dagger} (\mc{G} \sigma_x)^{-1} \hat{J}$ part of the action, where we neglect terms $O(\dot{\mb{R}}^3)$. The advantage of the Keldysh rotation \eqref{Rotation} is that the operator $(\mc{G} \sigma_x)^{-1}$ is identified with the Green's function of fluctuations
\begin{align}
(\mc{G} \sigma_x)^{-1} = \begin{pmatrix}
G^K&G^R\\
G^A&0
\end{pmatrix},
\end{align}
where $G^{R,A}=(i \bar{S} \sigma_z \partial_t \pm i0 - \mc{H} - V(\mb{r},t))^{-1}$ are the retarded and advanced Green's functions defined as
\begin{align}
G^{R} (t,t') &=-\frac{i}{\bar{S}} \sigma_z \Theta(t-t') T_{+} e^{-i \sigma_z \int_{t'}^{t} d\tilde{t} (\mc{H}+V(\mb{r},\tilde{t}))/\bar{S}} \nonumber \\
G^{A}	 (t,t') &= \frac{i}{\bar{S}} \sigma_z \Theta(t'-t) T_{-} e^{-i \sigma_z \int_{t'}^{t} d\tilde{t} (\mc{H}+V(\mb{r},\tilde{t}))/\bar{S}} 
\end{align}
where $T_{\pm}$ time orders in chronological/antichronological order. The Keldysh Green's function equals $G^K = G^R F - F G^A$, while function $F$ is known only at thermal equilibrium. In Fourier space and at thermal equlibrium, the Keldysh Green's function is $G^K(\omega) = (G^R(\omega) - G^A(\omega)) \coth(\beta \omega/2)$. The term under study takes the following form
\begin{align}
\mc{S}_1 = - N_A (J_q^{\dagger} \cdot G^R J_c + J_c^{\dagger} \cdot G^A J_q + J_q^{\dagger} \cdot G^K J_q) 
\end{align}
By minimizing $\mc{S}_1$ we arrive at
\begin{align}
\frac{\delta \mc{S}_1}{\delta R_q^{j}(t'')} =- \int_{t_0}^{t''} d t' \dot{R}_c^j(t')  \gamma_{ji} (t'',t') 
\end{align}
where $\gamma_{ji}(t'',t')$ is a dissipation kernel which originates from the coupling of the skyrmion with magnon modes
\begin{align}
\gamma_{ij}(t'',t')& = - N_A \bar{S}^2 \int d\mb{r} d\mb{r}' \left( f_j^{\dagger}(\mb{r}) \sigma_z \partial_{t''} G^{A}(\mb{r},\mb{r}', t',t'') \sigma_z f_i(\mb{r}')+f_i^{\dagger}(\mb{r}) \sigma_z \partial_{t''} G^{R}(\mb{r},\mb{r}', t'',t') \sigma_z f_j(\mb{r}') \right) \,.
\label{Damping0}
\end{align}
Note that the term $\sim J_q^{\dagger} \cdot G^K J_q$ gives rise to random forces which naturally emerge from the coupling of the skyrmion with the reservoir of magnon fluctuations. In addition, the semiclassical terms analyzed so far are of order $\mc{O}[N_A S]$, while quantum terms originating from the determinant appearing in \eqref{PartitionFunc} are of order $\mc{O}[(N_A S)^{0}]$ and are neglected under the assumption $N_A S \gg 1$. The Langevin equation for the skyrmion, including quantum dissipation under an external oscillating field is presented in a forthcoming work. Gathering all the terms, the skyrmion obeys the quasiclassical equation of motion
\begin{align}
\tilde{Q} \epsilon_{ij} \dot{R}^{j}(t) +   \int_{t_0}^{t} dt' ~ \dot{R}^j (t') \gamma_{ji}(t,t')  = \Fext^{i}(t) 
\label{EquationTime}
\end{align}
provided that $\tilde{Q} = 4 \pi N_A Q \bar{S}$ and that $\Fext^{i}(t) =- N_A \Theta(t-t_0) \int d\mb{r} ~\mb{b}(\mb{r},t) \cdot \partial_i \mb{m}_0 (\mb{r} -\mb{R}(t))$. Without loss of generality we assume the time of preparation of the initial state $t_0$ coincides with the onset of the external force. Here we make the replacement $R_c^j(t) \rightarrow R^j(t)$. 

\section{D. Effective Mass induced by spatial confinement}
In this section we study the effective dissipation induced by spatial confinement. For this reason we consider a skyrmion of radius $\lambda$ placed in a magnetic nanoribbon of width $4 \lambda$ in the $y$ direction and infinite in the $x$ direction. In such a linear track the repulsive forces imposed by surface twists create a potential of the form $\Wconf(y) = 2 \cosh( y h/2)$ \cite{Meynell14}. In addition, for the dissipation kernel we have $\gamma_{ij}(t,t')=\gamma^0_{ij}(t-t')$, where $\gamma^0_{ij}(t-t')$ is given by \eqref{Damping0}, while propagators equal $G^{R,A}= G_{0}^{R,A}=(i \bar{S} \sigma_z \partial_t \pm i0 - \mc{H}-\Vconf(y))^{-1}$. The magnon Hamiltonian acquires an additional potential term $\Vconf$ due to the confining potential experienced by the magnetic excitations. The equation of motion in frequency space is
\begin{align}
\tilde{Q} \varepsilon_{ij} (-i \omega) R^{j}(\omega) + (-i \omega) R^{j}(\omega) \gamma^0_{ji}(\omega) = \mc{K}_{ij} R^j(\omega) \,,
\label{FreqEq}
\end{align}
where $\tilde{Q}= 4 \pi N_A Q \bar{S}$ and where we define the sping constant $\mc{K}_{ij}= \delta_{ij} K^j$ which parametrizes the effects of the lateral parabolic confining potential $\Wconf(y)$ and $K^i = N_A \frac{\delta^2}{\delta R_i^2} \int d \mb{r} \Wconf(y-R_i) $. A time dependent function $g(t)$ is expanded in Fourier series as $g(t) = \int_{-\infty}^{\infty} \frac{d\omega}{2 \pi} g(\omega) e^{-i \omega t}$. The retarded and advanced Green's functions in Fourier space in the basis of eigenfunctions of $\mc{H}$ are given by
\begin{align}
G_0^{R,A}(\mb{r},\mb{r}', \omega) = \frac{1}{\bar{S}} \sum_{q=\pm 1}\sum_{n} \frac{ q \sigma_z \Psi^{q}_n(\mb{r}) (\Psi^{q}_n(\mb{r}'))^{\dagger}\sigma_z}{\omega \pm 0i -\bar{\varepsilon}^{q}_n}  \,.
\label{GreenFun0}
\end{align}
States $\Psi_n(\mb{r})$ are solutions of the eigenvalue problem $\mc{H} \Psi_n = \varepsilon_n \sigma_z \Psi_n$ and by definition $\bar{\varepsilon}_n = \varepsilon_n /\bar{S}$. Here index $q= \pm 1$ distinguish between particle states ($q=1$) with eigenfrequency $\bar{\varepsilon}_n^{1} =+ \bar{\varepsilon}_n$ and antiparticle states  ($q=-1$) with eigenfrequency $\bar{\varepsilon}_n^{-1} =- \bar{\varepsilon}_n$. Antiparticle states are obtained from particle states as $\Psi_n^{-1} = C \sigma_x \Psi_n^1$, where $C$ is the operator of complex conjugation. The damping kernel in Fourier space is equal to
\begin{align}
\gamma^0_{ji}(\omega) = N_A \bar{S} \sum_{q=\pm 1} \sum_n^{}{'} (- i \omega) \left( \frac{ q \mc{A}_{ji}^{n,q}}{\omega+ 0i+\bar{\varepsilon}^{q}_n}+\frac{ q \mc{A}_{ij}^{n,q}}{-\omega- 0i+\bar{\varepsilon}^{q}_n} \right) \,,
\label{gamma0}
\end{align}
where the prime in the sum denotes omission of the zero modes and where we define $\mc{A}_{ij}^{n,q} = ( f_i^{\dagger} \sigma_z \Vconf \Psi^{q}_n  ) ( (\Psi^{q}_n)^{\dagger} \Vconf \sigma_z f_j )/(\varepsilon_n^2)$, where we use the notation $(A) = \int d\mb{r} A$. From the structure of the matrix elements $\mc{A}_{ij}^{n,q}$ we can show that $\mbox{Im}[\mc{A}^{n,q}_{xx}]=0=\mbox{Im}[\mc{A}^{n,q}_{yy}]$, $\mbox{Re}[\mc{A}^{n,q}_{xy}]=0=\mbox{Re}[\mc{A}^{n,q}_{yx}]$, and $\mbox{Im}[\mc{A}^{n,q}_{xy}]=-\mbox{Im}[\mc{A}^{n,q}_{yx}]$. Further, $\mc{A}^{n,1}_{ii} = \mc{A}^{n,-1}_{ii}$, $\mc{A}^{n,1}_{xy} = -\mc{A}^{n,-1}_{xy}$ and $\mc{A}^{n,1}_{yx} = -\mc{A}^{n,-1}_{yx}$. Therefore, the diagonal part of the damping kernel $\gamma^0_{ij}(\omega)$ in frequency space is 
\begin{align}
\gamma^0_{ii}(\omega)= N_A \bar{S} \sum_{q =\pm 1} \sum_{n}^{}{'} (-i \omega) \frac{2 q  \mbox{Re}[\mc{A}^{n,q}_{ii}]\bar{\varepsilon}^{q}_n}{(\bar{\varepsilon}_{n}^{q})^2 - \omega^2 - 2 0 i \omega} \,.
\label{gammaw0x}
\end{align} 
We observe that the memory kernel has a power law form at low frequencies $\gamma_{ii}^{0}(\omega) \propto \omega^{s}$, with $s=2$ which corresponds to $\textit{super - Ohmic}$ friction. The power-law holds in the frequency range $0 \leq \omega \lesssim\egap/\bar{S}$, where $ \egap$ is the smallest magnon gap. Thus, the main effect of the environmental coupling on the asymptotic dynamics in an induction of an effective mass
\begin{align}
\mc{M}_{ii} = 2 N_A \bar{S} \sum_{q=\pm 1} \sum_{n}^{}{'} q \frac{\mbox{Re}[\mc{A}^{n,q}_{ii}] }{\bar{\varepsilon}^{q}_{n}} \,.
\label{MassStatic}
\end{align}
already predicted in \cite{Psaroudaki17}. The off diagonal part of the damping kernel equals
\begin{align}
\gamma^0_{xy}(\omega)= N_A \bar{S} \sum_{q =\pm 1} \sum_{n}^{}{'} (-i \omega)^2 \frac{2 q  \mbox{Im}[\mc{A}^{n,q}_{xy}]}{(\bar{\varepsilon}_{n}^{q})^2 - \omega^2 - 2 0 i \omega}\,,
\label{gammaw0y}
\end{align}
with a sub-leading spectal \textit{super-Ohmic} behavior at low frequencies with $s=3$, irrelevant for the skyrmion's asymptotic dynamics. Note that so far we have neglected any scattering mechanism through which the magnon environment can relax. 
\section{E. Driven skyrmion by an oscillating magnetic field}
We now assume that at an initial time $t_0$ we turn on an external nonuniform magnetic field  $\mb{b}(\mb{r},t)= \Theta(t-t_0) b(\mb{r}) \cos(\wext t)\hat{z}$ with an oscillating time dependence at a frequency $\wext$. The equation of motion in frequency space is given by
\begin{align}
\tilde{Q} \varepsilon_{ij}(- i \omega) R^{j}(\omega) +\int_{-\infty}^{\infty} \frac{d \omega'}{2 \pi} (-i \omega') R^{j}(\omega') \gamma_{ji}(-\omega',\omega) =\Fext^i(\omega)  +\mc{K}_{ij} R^j(\omega)\,.
\label{EqFreqSimple}
\end{align}
The magnon Green's function now acquires an additional term  $\mc{G}=  i \bar{S}\sigma_z \partial_{t} - \mathcal{H} - V(\mb{r},t)$ with $V(\mb{r},t) = \mb{b}(\mb{r},t) \cdot \mb{D} = V(\mb{r}) h_z \Theta(t-t_0)\cos(\wext t) $ and $ \mb{D} =  \delta_{\chi^\dagger}\delta_\chi \mb{m} |_{\chi=\chi^\dagger=0}$. The analysis is based on calculating the damping kernel of Eq.~\eqref{Damping0}, provided that the retarded and advanced propagator in the external potential are specified in first order perturbation theory
\begin{align}
G^{R,A}(t,t')= G_0^{R,A}(t,t')+ G_0^{R,A} \circ V \circ G^{R,A} \,,
\label{RAPropagators}
\end{align} 
with $G_0^{R,A}(t,t')$ given by Eq.\eqref{GreenFun0}. We introduce the compact notation 
\begin{align}
G \circ V \circ G \equiv \int d\mb{r}_1 dt_1 G(\mb{r},\mb{r}_1,t,t_1) V(\mb{r}_1,t_1)G(\mb{r}_1,\mb{r}',t_1,t')  \,.
\end{align}
In order to study the skrymion's propagation under resonant excitation when the external frequency $\wext$ is close to one of the eigenfrequencies $\wres=\varepsilon_{n}/\bar{S}$ we need to include a relaxation rate $\Gamma$, {\it i.e.}  $\varepsilon_{n} \rightarrow \varepsilon_n - i \Gamma$. This term phenomenologically parametrizes any mechanism that could lead to relaxation of the magnon bath dynamics, for example the back-action of the skyrmion to the bath, magnon-magnon and phonon-magnon interaction. Since the external field is periodic in time we can use an exact expression of the form 
\begin{align}
\gamma_{ji}(t,t') = \gamma_{ji}(t,t-t')= \gamma^0_{ji}(t-t') + \sum_{k}\gamma^{k}_{ji}(t-t') e^{ik \wext t} \,.
\end{align}
This suggests that time differences $t-t'$ describe memory effects for the skyrmion's path under the drive of the field at time $t$. 
The dissipation term in Fourier space is simplified as follows,
\begin{align} 
&\int_{-\infty}^{\infty} \frac{d \omega'}{2 \pi} (-i \omega') R^{j}(\omega') \gamma_{ji}(-\omega',\omega)= (-i \omega) \gamma_{ji}(t,\omega) R^{j}(\omega)
\label{Simpl}
\end{align}
where $\gamma_{ji}(t,\omega) =  \gamma^0_{ji}(\omega) +\Delta\gamma_{ji}(t,\omega)$. We find that 
\begin{align}
\Delta\gamma_{ji}(t,\omega) = \left( e^{-i\wext t} \Delta\bar{\gamma}_{ji}(\omega,-\wext)+e^{i \wext t} \Delta\bar{\gamma}_{ji}(\omega,\wext) \right)
\end{align}
where
\begin{align}
\Delta \bar{\gamma}_{yx}(\omega,\pm \wext) =h_z N_A \sum_{\substack{n,n' \\ q,q'=\pm 1}}^{}{'} \mc{B}_{yx}^{nq,n'q'} \frac{ g^{\pm}(\omega) (\pm i \bar{\varepsilon}_{n'} \wext + (\Gamma-i \omega) (\bar{\varepsilon}_{n'} +\bar{\varepsilon}_n))}{\Delta(\omega)}  \,,
\end{align}
and
\begin{align}
\Delta \bar{\gamma}_{ii}(\omega,\pm \wext) =h_z N_A \sum_{\substack{n,n' \\ q,q'=\pm 1}}^{}{'}\mc{B}_{ii}^{nq,n'q'}  \frac{g^{\pm}(\omega)( i \bar{\varepsilon}_{n'}\bar{\varepsilon}_{n}  + (\Gamma-i \omega) (\pm \wext -i \Gamma - \omega))}{\Delta(\omega)} \,.
\end{align}
where $g^{\pm} (\omega) = (-\omega \pm \wext)(\bar{\varepsilon}_{n} \pm \wext +i \Gamma+\omega)(-\bar{\varepsilon}_{n} \pm \wext +i \Gamma+\omega)$ and $\Delta(\omega) =(\bar{\varepsilon}_{n'}^2+(\Gamma-i \omega)^2)((\bar{\varepsilon}_{n}+\wext)^2-(i \Gamma+\omega)^2)((\bar{\varepsilon}_{n}-\wext)^2-(i \Gamma+\omega)^2) $. 
Here we define $\bar{\varepsilon}_n=\varepsilon_n/\bar{S}$ and $\mc{B}_{ij}^{nq,n'q'} = q q'( f_i^{\dagger} \sigma_z \Wconf \Psi^{q}_n)((\Psi^q_n)^{\dagger}V\Psi^{q'}_{n'})((\Psi^{q'}_{n'})^{\dagger} \Wconf \sigma_z f_j )/(\varepsilon_n\varepsilon_{n'})$. As long as we are interested in asymptotic times we can focus on the low frequency power law behavior of the friction kernel $\gamma_{ji}(t,\omega)$, {\it i.e.} perform a Taylor expansion around the origin as
\begin{align}
\gamma_{ji}(t,\omega)& \simeq \gamma_{ji}(t,0) +(- i \omega) \pt_{\omega} \gamma_{ji}(t,\omega)\vert_{\omega =0}  +\mc{O}(\omega^2) \,,
\label{Approximation}
\end{align}
which holds up to frequencies $\omega \ll \Gamma/\bar{S}$ and $\omega \ll \egap/\bar{S}$. We note that the term $\gamma_{ji}(t,0)$ corresponds to a time-dependent Ohmic term proportional to skyrmion's velocity in real time. Therefore, there is an \textit{super-Ohmic} to \textit{Ohmic} crossover behavior for both the diagonal as well as the off diagonal damping. Upon substituting the approximation of Eq.~\eqref{Approximation} in Eqs.~\eqref{FreqEq}-\eqref{Simpl} and performing an inverse Fourier transform we arrive at a local equation of motion, 
\begin{align}
 \begin{bmatrix}
D_x(t) + M_x(t) \pt_t& Q(t) + G(t) \pt_t\\
-Q(t) -G(t)\pt_t & D_y(t) + M_y(t) \pt_t
\end{bmatrix} \binom{\dot{R}_x}{\dot{R}_y}  = \binom{\Fext^x(t)}{K^y R_y} \,,
\label{EquationReduced}
\end{align}
where $D_i (t) = \delta D_i \sin(\wext t)$, $G(t) = \delta G \sin(\wext t)$, $M_{j}(t) =  \mc{M}_{jj}+ \delta M_j \cos(\wext t)$ and $Q(t)= \tilde{Q}_0 +\delta Q \cos(\wext t) $.  For $\Gamma \ll \egap$, the time dependent dissipation terms are found by the following closed formulas
\begin{align}
\delta D_i&= 2 N_A h_z \wext\sum_{\substack{n,n' \\ q,q'=\pm 1}}^{}{'} \frac{\mc{B}_{ii}^{nq,n'q'} \bar{\varepsilon}_n^{q}}{\bar{\varepsilon}_{n'}^{q'}((\bar{\varepsilon}_n^{q})^2-\wext^2)} + \mc{O} (\Gamma)  \,,\\
\delta Q&=- 2 i N_A h_z \wext^2\sum_{\substack{n,n' \\ q,q'=\pm 1}}^{}{'} \frac{\mc{B}_{yx}^{nq,n'q'}}{\bar{\varepsilon}_{n'}^{q'}((\bar{\varepsilon}_n^{q})^2-\wext^2)}+ \mc{O} (\Gamma) \,,
\end{align}
and 
\begin{align}
\delta M_{i}(t)&= -2 N_A h_z \sum_{\substack{n,n' \\ q,q'=\pm 1}}^{}{'} \frac{\mc{B}_{ii}^{nq,n'q'} (\bar{\varepsilon}_{n'}^{q'} (\bar{\varepsilon}_{n}^{q})^3 +\bar{\varepsilon}_{n}^{q}(\bar{\varepsilon}_{n'}^{q'}+\bar{\varepsilon}_{n}^{q})\wext^2-\wext^4)}{\bar{\varepsilon}_{n'}^{q'}((\bar{\varepsilon}_n^{q})^2-\wext^2)}+ \mc{O} (\Gamma)  \,,\\
\delta G(t)&=- 2 i N_A h_z \wext \sum_{\substack{n,n' \\ q,q'=\pm 1}}^{}{'} \frac{\mc{B}_{yx}^{nq,n'q'}\bar{\varepsilon}_{n}^{q}(2 \bar{\varepsilon}_{n'}^{q'}\bar{\varepsilon}_{n}^{q} +(\bar{\varepsilon}_{n}^{q})^2-\wext^2)}{\bar{\varepsilon}_{n'}^{q'}((\bar{\varepsilon}_n^{q})^2-\wext^2)} + \mc{O} (\Gamma) \,.
\end{align} 
\subsection{F. Skyrmion as Saddle Point Configuration}
In this section we give the explicit parametrization of the skyrmion by finding the approximate saddle point configuration of the following free energy functional
\begin{align}
\mc{F}(\Phi , \Pi)= & (\nabla \Theta)^2+ \sin^2\Theta( \nabla \Phi)^2 - h \cos\Theta  \nonumber \\
&+ \left(\cos(\phi-\Phi)\frac{\pt \Theta}{\pt \rho} - \sin\Theta\cos\Theta \sin(\phi-\Phi)\frac{\pt \Phi}{\pt \rho} - \frac{1}{\rho} \sin(\phi -\Phi)\frac{\pt \Theta}{\pt \phi} -\frac{1}{\rho}\sin\Theta\cos\Theta\cos(\phi-\Phi)\frac{\pt \Phi}{\pt \phi} \right) \,.
\label{EnergyDensity}
\end{align}
In the following we consider the regime of isolated skyrmions as a metastable state of the ferromagnetic background $\mb{m}=(0,0,1)$. Rotationally symmetric solutions are described by
\begin{equation}
\Theta_0(\rho,\phi)=\Theta_0(\rho) , \qquad \Phi_0(\rho,\phi)= \phi + \pi/2 \,,
\label{SolTheta}
\end{equation}
for a skyrmion with topological number $Q=-1$. The skyrmion energy with respect to the uniform state is, 
\begin{align}
\mc{F}_0(\Phi_0,\Pi_0)= (\nabla \Theta_0)^2 +(\frac{1}{\rho^2}+\kappa) \sin ^2\Theta_0 + h(1-\cos \Theta_0) +\Theta_0' +\frac{1}{\rho}\sin \Theta_0 \cos \Theta_0 \,.
\end{align}
The structure of the stationary skyrmion is determined by the Euler--Lagrange equation,
\begin{align}
\Theta_0''(\rho)+\frac{\Theta_0'(\rho)}{\rho}-\frac{\sin\Theta_0 \cos\Theta_0 }{\rho^2} + \frac{\sin^2\Theta_0}{\rho} -\frac{h}{2} \sin\Theta_0 =0\,,
\label{EulerEq}
\end{align}
with boundary conditions $\Theta_0(0)=\pi$ and $\Theta_0(\infty)=0$. Because there is no known analytic solution, in the large-radius limit the following function can be used as an approximate solution
\begin{equation}
\Theta_0(\rho) = A \cos^{-1}\left(\tanh (\frac{\rho - \lambda}{\Delta_0}) \right)\,,
\label{Profile_Lrg}
\end{equation} 
where $A=\pi/\cos^{-1}(\tanh(-\lambda/\Delta_0))$. The parameters $\lambda$ and $\Delta_0$ are calculated numerically by fitting the approximate function \eqref{Profile_Lrg} to the numerical solution of the Euler-Lagrange equation \eqref{EulerEq}. 

\subsection{G. Magnon Spectrum}
\label{sec:ap2}

In this section we discuss the magnon spectrum over the skyrmionic background. Magnon modes are solutions of the eigenvalue problem $\mc{H} \Psi_{n} = \varepsilon_{n} \sigma_z \Psi_{n}$ (EVP), where the explicit form of the Hermitian Hamiltonian $\mc{H}$ is given in Eq.~\eqref{EVPTransfm} below. The Hamiltonian is invariant under $C \sigma_x \mc{H} (C\sigma_x)^{\dagger} = \mc{H}$, with $C$ the operator of complex conjugation. Therefore an additional class of solutions is generated with negative eigenfrequencies. More specifically, particle states $\Psi^{1}_{n}$ are solutions of the eigenvalue problem (EVP) with eigenfrequencies $\varepsilon_n \geq 0$ and antiparticle states $\Psi_n^{-1}= C\sigma_x \Psi^1_{n} $ are solutions of the same EVP with eigenfrequency $-\varepsilon_{n}$ \cite{Sheka04}. The biorthogonality conditions for the solutions $\Psi^{q=\pm1}_{n}$ are
\begin{align}
\langle \Psi^{q}_{n} \vert\sigma_z \vert \Psi^{q'}_{m} \rangle &= q \delta_{q,q'} \delta_{n,m} \,.\nonumber \\
\label{Orth}
\end{align}
Similarly, the unity operator is given by $\mathbb{1} = \sum_{q=\pm 1} \sum_{n} q \vert \Psi^{q}_{n} \rangle \langle \Psi^{q}_{n} \vert \sigma_z$ and the trace of an operator is $\mbox{Tr}(A) = \sum_{q=\pm 1} \sum_{n} q  \langle \Psi^{q}_{n} \vert \sigma_z  A \vert \Psi^{q}_{n} \rangle$. For reasons of convenience, in this section we use the notation $\Psi_n$ to denote particle states, and antiparticles states are recovered using the particle-antiparticle symmetry described above. Magnon scattering states are obtained for energies $\varepsilon_n \geq \varepsilon_{\mbox{\tiny{MS}}}$, with $\varepsilon_{\mbox{\tiny{MS}}}=h$. In addition to scattering states, we expect localized modes that correspond to deformations of the skyrmion into polygons (breathing modes) in the range $0 < \varepsilon_n < \varepsilon_{\mbox{\tiny{MS}}}$. To begin with, we seek for solutions of the EVP where the Hamiltonian  is given by
\begin{equation}
\mc{H}=2[- \nabla^2 + U_0(\rho)] \mathds{1} + 2 W(\rho) \sigma_x - 2 i V(\rho) \frac{\pt}{\pt \phi}\sigma_x \,.
\label{EVPTransf}
\end{equation}
For the potential terms we have $V(\rho)=\frac{2\cos \Theta_0}{\rho^2} -\frac{ \sin \Theta_0}{\rho}$, 
\begin{align}
W(\rho)&=\frac{\sin 2\Theta_0 }{4\rho}- \frac{(\Theta_0')^2}{2}+ \frac{1}{2 \rho^2}\sin^2 \Theta_0 - \frac{ \Theta_0'}{2} \,,
\end{align}
and 
\begin{align}
U_0(\rho)= \frac{h \cos \Theta_0}{2}  -\frac{3\sin 2 \Theta_0 }{4 \rho}   -\frac{(\Theta_0')^2}{2}+\frac{1}{4 \rho^2} (1+ 3 \cos 2\Theta_0)-\frac{\Theta_0'}{2} \,, 
\end{align}
Next, we represent solutions in terms of wave expansions $\Psi_{n}=e^{i  m \phi}\psi_{n,m}(\rho)$, and the EVP is written as $\mc{H}_m  \psi_{n,m}(\rho) = \varepsilon_{n,m} \sigma_z  \psi_{n,m}(\rho)$ with 
\begin{equation}
\mc{H}_m =  2(- \nabla^2_{\rho}+ U_0(\rho) +\frac{ m^2}{\rho^2} ) \mathds{1}  
+ 2V(\rho) m \sigma_z + 2W(\rho) \sigma_x\,,
\label{EVPTransfm}
\end{equation}
where $\nabla^2_{\rho}= \frac{\pt^2}{\pt_{\rho}}+\frac{1}{\rho} \frac{\pt}{\pt_{\rho}}$. These eigenfunctions are normalized such that
\begin{equation}
\int_0^{\infty} d \rho~ \rho \psi_{n,m}^{\dagger} (\rho)\sigma_z \psi_{n',m}(\rho) =\delta_{n,n'} \,.
\label{Normaliz1}
\end{equation}
One of the translational modes with zero energy is given by 
\begin{equation}
\Psi^{1}_1= e^{i\phi} \frac{1}{\sqrt{8}}\binom{\Theta_0' -\frac{1}{\rho} \sin \Theta_0}{\Theta_0' +\frac{1}{\rho} \sin \Theta_0}\,
\label{ZMode1}
\end{equation}
and the other is found using the particle-antiparticle symmetry
\begin{equation}
\Psi^{-1}_{1}= C \sigma_x \Psi^{1}_1 = e^{-i\phi} \frac{1}{\sqrt{8}}\binom{\Theta_0' +\frac{1}{\rho} \sin \Theta_0}{\Theta_0' -\frac{1}{\rho} \sin \Theta_0} \,.
\label{ZMode2}
\end{equation}
Here, the upper index corresponds to particle/antiparticle while the lower to the quantum number $m$. We focus on the calculation of the localized modes $\Pbs_m$ with energy $\varepsilon_m$ which are found variationally using the trial functions 
\begin{align}
\Pbs_m(\mb{r}) = \frac{1}{\sqrt{2 \pi}} e^{i m \phi} \binom{a_m f_0(\rho)}{b_m f_0(\rho)}\,,
\label{BS_LL}
\end{align}
where $f_0(\rho)=A \rho /\cosh(\frac{\rho-\lambda}{\Delta_0})$ and $A$ is chosen such that $\int d\rho \rho f_0^2 =1$. The condition of minimizing the energy functional $\mc{U}=\int d\mb{r} (\Pbs_m)^{\dagger} (\mc{H}_m - \sigma_z \varepsilon_m) \Pbs_m $, along with normalization conditions satisfied by the functions $\Pbs_m(\rho)$ will specify the eigenenergies $\varepsilon_m$ as well as the variational parameters $a_m$ and $b_m$. For $h = 0.5$ ($\lambda=3.3$ and $\Delta_0 = 1.2$) we find the following four bound states, $\varepsilon_2= 0.0704$, $\varepsilon_0=0.1358$, $\varepsilon_3= 0.3358$ and $\varepsilon_{-1}= 0.4498$, lying below the energy gap of the scattering states $\ems =h =0.5$.

\end{document}